\documentclass[twocolumn,aps,superscriptaddress,prl,showpacs,floatfix]{revtex4}

\usepackage{graphicx,amsmath,amssymb}
\usepackage{bm}

\newcommand{\cM}{{\cal M}}
\newcommand{\pa}{\partial}
\newcommand{\pr}{p_p(x)}
\newcommand{\lik}{p_l(x)}
\newcommand{\llik}{\ln\lik}
\newcommand{\blik}{p_l^{\beta}(x)}
\newcommand{\lan}{\langle}
\newcommand{\ran}{\rangle}

\begin{document}

\title{Prior-predictive value from fast growth simulations} 

\author{H. Ahlers}
\email{ahlers@theorie.physik.uni-oldenburg.de}
\author{A. Engel}
\email{engel@theorie.physik.uni-oldenburg.de} 

\affiliation{Institut f\"ur Physik, Carl-von-Ossietzky-Universtit\"at,
     26111 Oldenburg, Germany }


\begin{abstract}
  Building on a variant of the Jarzynski equation we propose a new
  method to numerically determine the prior-predictive value in a
  Bayesian inference problem. The method generalizes thermodynamic
  integration and is not hampered by equilibration problems. We
  demonstrate its operation by applying it to two simple examples and
  elucidate its performance. In the case of multi-modal posterior
  distributions the performance is superior to thermodynamic
  integration. 
\end{abstract}

\maketitle

\section{Introduction}

Bayesian methods of inference \cite{Jaynes,GCSR,LeHs} are
playing an ever growing role in the statistical analysis of data in
physics and other natural sciences \cite{BBDS,Agostini,Doserev}. Among
its particular virtues is the ability to perform model selection,
i.e. to quantitatively assess the appropriateness of a
particular model irrespective of concrete parameter values. This is
accomplished by calculating what is called the {\it evidence} or the
{\it prior-predictive value}.   

Building on rather general and essentially simple principles the
efficiency of Bayesian methods in practical applications depends
crucially on the implemented numerical algorithms. A major difficulty 
common to Bayesian data analysis is the calculation of integrals in
high-dimensional spaces which are dominated by contributions from
small and labyrinthine regions. Similar problems are typical for the
numerical determination of the partition function in statistical
mechanics. It is therefore no surprise that some of the tools
developed in statistical mechanics have found their way into the
arsenal of methods used in Bayesian inference. 

In the present paper we show that recent progress in the
statistical mechanics of non-equilibrium processes
\cite{Jar,JarFT,Crooks,Seifert} entails new possibilities to estimate
the prior-predictive value and to average with the
posterior distribution of a Bayesian analysis. The new method relies
on the Monte-Carlo (MC) simulation of a {\it non-stationary} Markov
process and is intermediate between straight MC sampling and
thermodynamic integration \cite{thdint}. It is generally superior to
the first and may also outperform the second. We first give a general
theoretical discussion and then analyze numerically two simple
examples. One of these was used already in \cite{LPD} to scrutinize
the efficiency of thermodynamic integration. The second is a
generalization thereof employing a bimodal likelihood.

\section{Theory}

We consider a standard Bayesian inference problem in which parameters
$x$ of a model $\cM$ are to be determined from data $d$. The prior
information about $x$ is coded in a prior distribution $p_p(x|\cM)$,
the likelihood of the data given certain values of $x$ is denoted by
$p_l(d|x,\cM)$. By Bayes' theorem the posterior distribution is given
by 
\begin{equation}
  \label{eq:postP}
  p_{\mathrm{post}}(x|d,\cM)=\frac{p_l(d|x,\cM)\,
    p_p(x|\cM)}{P(d|\cM)}\, .
\end{equation}
Our central quantity of interest is the normalization of the 
posterior, the so called evidence or prior-predictive value, defined
by  
\begin{equation}
  \label{eq:defppv}
  P(d|\cM)=\int dx \, p_l(d|x,\cM)\, p_p(x|\cM)\, .
\end{equation}
It quantifies the likeliness of the data for the particular model
$\cM$ under consideration and is therefore crucial for the comparison
between different models. 
For the following manipulations the dependence of $p_p(x|\cM)$ and
$p_l(d|x,\cM)$ on $x$ will be the important one, in order to lighten the
notation we will therefore suppress the dependencies on $d$ and $\cM$
in these quantities. Generically the integral in (\ref{eq:defppv}) is
dominated by intricately shaped regions in a high-dimensional space
and is therefore difficult to determine. 

A possible remedy for this problem is motivated by the method of
thermodynamic integration used in statistical mechanics. To this end
one introduces the auxiliary quantity
\begin{equation}
  \label{eq:defZ}
  Z(\beta):=\int dx \, \big(\lik\big)^\beta \pr\, .
\end{equation}
Clearly $Z(0)=1$ due to the normalization of the prior
distribution and $Z(1)=P(d|\cM)$, the desired quantity. Moreover 
\begin{align}\nonumber
  \frac{\pa}{\pa \beta}\ln Z(\beta)&=\frac{1}{Z(\beta)}\int dx \llik
  \;\blik \,\pr\\ 
  &=:\lan \llik\ran_\beta \, , \nonumber
\end{align}
where the $\beta$-average in the last line is taken with the
distribution 
\begin{equation}
  \label{eq:defPbeta}
   P_\beta(x):=\frac{1}{Z(\beta)}\; \blik\,\pr \, . 
\end{equation}
We therefore get 
\begin{equation}\label{eq:thdint}
  \ln P(d|\cM)=\int_0^1 d\beta\, \frac{\pa}{\pa \beta}\ln Z(\beta)
        =\int_0^1 d\beta\, \lan \llik\ran_\beta \, ,
\end{equation}
from which the name thermodynamic integration for the method becomes
clear. 

Since averages are much more efficiently calculated from 
MC simulations than normalization factors (\ref{eq:thdint}) offers a
convenient way to determine $P(d|\cM)$ from equilibrium simulations for
just a few values of $\beta$ between 0 and 1. The method was suggested
in \cite{thdint}, its advantages over a straight-forward MC estimation
of $P(d|\cM)$ from (\ref{eq:defppv}) was demonstrated for a simple
test example in \cite{LPD}.   

Let us consider the MC simulations necessary to implement
(\ref{eq:thdint}) in somewhat more detail. We first discretize the
$\beta$-interval by introducing $M$ values $\beta_m, m=1,...,M$, with 
$0=\beta_1<\beta_2<...<\beta_M=1$. For each $\beta_m$ we generate a
trajectory $x_t$ with a discrete time $t=1,2,...$ measured in MC
steps. These trajectories are realizations of a Markov process 
with transition probability $\rho(x,x';\beta_m)$ which leaves the
distribution defined in (\ref{eq:defPbeta}) invariant,
i.e. which satisfies 
\begin{equation}
  \label{eq:inv}
  P_{\beta_m}(x)=\int dx' \rho(x,x';\beta_m) P_{\beta_m}(x')\, .
\end{equation}
A sufficiently long Markov chain is now generated for each $\beta_m$
in order to get a reliable estimate for $\lan \llik\ran_\beta$ to be
used in (\ref{eq:thdint}). This equilibration of the system at each
value of $\beta$ is the main bottleneck of the method. In realistic
situations with a multi-modal or otherwise complicated structure of
the likelihood it may become very slow and special care must be taken
which elements of the trajectory $x_t$ to use for the estimation of
the average  $\lan \llik\ran_\beta$. As a rule specific quantities need to be
identified and monitored which indicate when the system
has approximately equilibrated. 

These equilibration problems may be avoided by building on recent
progress in the statistical mechanics of non-equilibrium processes 
\cite{Jar,JarFT,Crooks,Seifert,LeSp,Chatelain}. In the present context it
gives rise to the following procedure to determine $Z(1)$. We fix a
time interval $t=1,...,N$ and generate a set of trajectories $x_t$
from a {\em non-stationary} Markov process such that for 
each trajectory $\beta$ changes from 0 to 1. More precisely we fix 
$M\leq N$ intermediate time points $t_m, m=1,...,M$ with
$1\leq t_1<t_2<....<t_M\leq N$ at which $\beta$ changes by $\Delta
\beta_m=\beta_{m+1}-\beta_m$. We call the set $\{\Delta\beta_m, t_m\}$
of these time points and the corresponding increments in $\beta$ the
{\em protocol} of the procedure.  

Consider a trajectory $\{x_t\}$ that starts in $x_1$ drawn from
the prior distribution $p_p$ and then evolves according to
$\rho(x,x';\beta_m)$ with $\beta_m$ fixed by 
the protocol $\{\Delta\beta_m, t_m\}$. The probability
$\mathcal{P}(\{x_t\})$ of the whole trajectory is given by
\begin{equation}
  \label{eq:probtraj}
  \mathcal{P}(\{x_t\})=p_p(x_1)
    \prod_{t=1}^{N-1} \rho(x_{t+1},x_t;\beta_m)\, .
\end{equation}
Consider now the trajectory dependent functional
\begin{equation}
  \label{eq:defR}
  R(\{x_t\})=\sum_{m=1}^{M-1} \Delta\beta_m \,\ln p_l(x_{t_m})\,  ,
\end{equation}
which is a random quantity due to its dependence on $\{x_t\}$. We will
show that its exponential average
\begin{align}\nonumber
  \langle e^{R}\rangle =&\nonumber
     \int\prod_{t=1}^N dx_t\; \mathcal{P}(\{x_t\})\; e^{R(\{x_t\})} \\
    =&\int\!\!\prod_{t=1}^N dx_t\;p_p(x_1)\nonumber
     \!\!\prod_{t=1}^{N-1} \rho(x_{t+1},x_t;\beta_m)
     \!\!\prod_{m=1}^{M-1}\Big(p_l(x_{t_m})\Big)^{\Delta\beta_m}
\end{align}
is equal to the desired quantity $Z(1)=P(d|\cM)$. 

To this end we first note that the integrations over the first $x_t$
with $1\le t<t_1$ are easily performed since during these time steps
$\beta=\beta_1=0$. Using (\ref{eq:inv}) repeatedly for $\beta=0$ we find
\begin{equation}
  \int\prod_{t=1}^{t_1-1} dx_t\;p_p(x_1)
    \prod_{t=1}^{t_1-1} \rho(x_{t+1},x_t;0)=p_p(x_{t_1})\, .
\end{equation}
Together with the $m=1$ term in (\ref{eq:defR}) and using
(\ref{eq:defPbeta}) as well as $\Delta\beta_1=\beta_2$ we hence obtain
\begin{align}\nonumber
  \int\prod_{t=1}^{t_1-1}& dx_t\;p_p(x_1)
    \prod_{t=1}^{t_1-1} \rho(x_{t+1},x_t;0) 
       \exp\big(\Delta\beta_1\,\ln p_l(x_{t_1})\big)\\\nonumber
  &=p_p(x_{t_1})\,p_l^{\beta_2}(x_{t_1})\\\nonumber
   &=Z(\beta_2)\, P_{\beta_2}(x_{t_1})\,.
\end{align}
The integrations over the $x_t$ with $t_1\le t<t_2$ can now be 
performed analogously. According to (\ref{eq:inv}) we get at first 
\begin{equation}
  \int\prod_{t=t_1}^{t_2-1} dx_t\;P_{\beta_2}(x_{t_1})
 \prod_{t=t_1}^{t_2-1} \rho(x_{t+1},x_t;\beta_2)=P_{\beta_2}(x_{t_2})\, . 
\end{equation}
since $\beta=\beta_2$ for the whole interval. Together with the second
term of the sum in (\ref{eq:defR}) we hence have 
\begin{align}\nonumber
  Z(\beta_2)&\,P_{\beta_2}(x_{t_2})\,
          \exp\big(\Delta \beta_2 \ln p_l(x_{t_2})\big)\\\nonumber
  &=p_p(x_{t_2})\;p_l^{\beta_2}(x_{t_2})\;
        p_l^{\beta_3-\beta_2}(x_{t_2})\\\nonumber
  &=p_p(x_{t_2})\;p_l^{\beta_3}(x_{t_2})\\\nonumber
  &=Z(\beta_3)\,P_{\beta_3}(x_{t_2})\, .
\end{align}
Iterating this procedure we finally arrive at 
\begin{equation}
  \label{eq:res}
  \lan e^{R}\ran=Z(\beta_M) \int dx_N P_{\beta_M}(x_N)=Z(1)
     =P(d|\cM)\, .
\end{equation}
Generalizing this relation to continuous protocols $\beta(t)$ with
$0\leq t\leq t_f$ we obtain 
\begin{equation}
  \label{eq:rescont}
 P(d|\cM)=\left\lan \exp\big(\int_0^{t_f} dt \ln p_l(x(t))\; 
       \frac{\pa}{\pa t} \beta(t)\big)\right\ran \, .
\end{equation}
Eqs. (\ref{eq:res}) and (\ref{eq:rescont}) are our central result. The
prior-predictive value can be determined from an 
exponential average of the quantity $R(\{x_t\})$ defined in
(\ref{eq:defR}) over an ensemble of MC trajectories $x_t$ generated
with a transition probability $\rho(x,x';\beta(t))$ that depends
explicitly on time via the protocol $\beta(t)$. Note that this protocol
must be the same for all trajectories $\{x_t\}$ that are used to
determine the average $\langle e^R \rangle$. It is nevertheless very
remarkable that the results (\ref{eq:res}) and (\ref{eq:rescont})
respectively do not depend on the details of this protocol. 

As a small aside we note that in a Bayesian analysis one is usually
more interested in {\it averages} with the posterior distribution
(\ref{eq:postP}) than in this distribution itself. By a slight
generalization of the above proof one can show for the
posterior average of some function $f(x)$ 
\begin{align}\nonumber
  \langle f\rangle_{\mathrm{post}}&=
   \int dx f(x) p_{\mathrm{post}}(x)=
   \int dx_N f(x_N) P_{\beta=1}(x_N)\\\label{eq:crooks}
   &=\frac{\langle e^R f(x_N)\rangle}{\langle e^R \rangle}\, ,
\end{align}
where the averages in the last line are taken with
$\mathcal{P}(\{x_t\})$. Posterior averages may hence be determined
from path averages starting from the prior distribution and
incorporating the weight factor $e^{R}$, cf. \cite{Crooks} for an
analogous result in the statistical mechanics framework. 

Two limiting cases of Eqs. (\ref{eq:res}) and (\ref{eq:rescont}) are of
interest. For $1 \ll M$ the system is manipulated in quasi-equilibrium
and $\beta$ and thus $P_{\beta}(x)$ change very slowly. Accordingly 
the Markov chain will explore much of the state space for a given small
$\beta$-interval and we may therefore replace $\ln p_l(x_{t_m})$ in
(\ref{eq:defR}) by $\langle \ln p_l(x)\rangle_{\beta_m}$. As a consequence
$R$ no longer depends on the individual realizations of the trajectories
$\{x_t\}$ and the average in (\ref{eq:rescont}) becomes superfluous. Therefore
\begin{align}\nonumber
P(d|\cM)&=\exp\big(\int_0^{t_f} dt\; \lan \ln p_l(x)\ran_\beta \; 
       \frac{\pa}{\pa t} \beta(t)\big)\\\label{eq:limittdint}
        &=\exp\big(\int_0^1 d\beta \lan \ln p_l(x)\ran_\beta\big)
\end{align}
which brings us back to thermodynamic integration, cf. (\ref{eq:thdint}).

In the opposite limit $\beta$ is changed in a single step from zero to
one at some time $t=t_*$, i.e. $\beta=\theta(t-t_*)$ with the
Heaviside $\theta$-function being 1 for positive arguments and zero
else. In this case the time integral in (\ref{eq:rescont}) picks up 
contributions from $t=t_*$ only and the average over the trajectories
$\{x_t\}$ reduces to the average over $x_*=x(t_*)$ with the prior
distribution $p_p$. We then find from (\ref{eq:rescont}) 
\begin{equation}
  \label{eq:thdpert}
P(d|\cM)=\int dx_* \exp(\ln p_l(x_*)) \, p_p(x_*)\, ,
\end{equation}
which is identical with (\ref{eq:defppv}). This limit is equivalent to 
what is called {\em thermodynamic perturbation} in statistical
mechanics \cite{Zwanzig}.  

The proposed method hence interpolates between the two extreme
variants (\ref{eq:defppv}) and (\ref{eq:thdint}) for the determination
of the prior-predictive value $P(d|\cM)$. It should be superior to the
straight application of (\ref{eq:defppv}) since the average in
(\ref{eq:rescont}) is already shortly after the start of the
trajectories influenced by the likelihood $p_l(x)$ which is, as a
rule, much sharper than the prior. It may outperform thermodynamic
integration (\ref{eq:thdint}) since no time-consuming equilibrations 
are necessary.  On the other hand, the exponential average in
(\ref{eq:rescont}) is known to be subtle. It is biased for finite sample sizes
\cite{Fox,GRB,Hummer} and dominated by rare realizations with very
big values of $R$  for processes too far from equilibrium \cite{Jarzynski2}.
Nevertheless, a strong argument in favour of the method is its applicability 
to systems that are hard to equilibrate and its great flexibility
to optimization for which the whole protocol $\beta(t)$ is at our disposal.

It is also worthwhile to mention that in many situations of interest
the probability distribution of $R$ is Gaussian with average $\langle
R\rangle$ and variance $\sigma_R^2$ \cite{SpSe,PaSch}. In this case
the average $\langle e^R \rangle$ can be calculated exactly with the
result 
\begin{equation}
  \label{eq:cum}
  P(d|\cM)=\exp\Big(\langle R\rangle+\frac{\sigma_R^2}{2}\Big)\, .
\end{equation}
The determination of $\langle R\rangle$ and $\sigma_R^2$ from the MC
simulations is, of course, much less demanding than the extraction of
the complete distribution of $R$. For general distributions 
(\ref{eq:cum}) gives just the first two terms of the cumulant
expansion.

\section{Two simple examples}

We now numerically investigate the performance of the method by
applying it to two simple, exactly solvable examples. The first one
employs a unimodal likelihood the second one uses a bimodal
one. These model systems are variants of those discussed in
\cite{LPD} and \cite{DKEL} respectively. The underlying inference
problem is to estimate an $n$-dimensional vector $x$ of parameters
from an $n$-dimensional vector $d$ of data with $n=128$. 

For the unimodal case both prior $ p_p(x)$ and likelihood $ p_l^u(x)$
are taken to be Gaussians with zero mean and variances  $\sigma_p^2$
and $\sigma_l^2$ respectively.  
\begin{align}
 \label{example-plpp}
 p_l^u(x)&=(2\pi \sigma_l^2)^{-n/2} \; 
 \exp(-\frac{\Vert d - x \Vert^2}{2\sigma_l^2})
 \\
 p_p(x)&=(2\pi \sigma_p^2)^{-n/2} \; 
 \exp(-\frac{\Vert x \Vert^2}{2\sigma_p^2})
\end{align} 
We will use $\sigma_p=10$ and $\sigma_l=1$ to model the typical
situation in which the prior is substantially broader than the
likelihood. Moreover we choose a data vector $d$ with 
\begin{equation}
  \label{eq:defd}
  \bar d^2=\frac{1}{n}\sum_{i=1}^n d_i^2=100\; ,
\end{equation}
which on the one hand ensures that the data are far from the center of
the prior $p_p(x)$, and on the other hand fixes the signal-to-noise
ratio SNR$=\sqrt{\bar d^2/\sigma_l^2}$ to be 10 as in \cite{LPD}. The
performance of the algorithms to be discussed below does not depend on
the particular values of the $d_i$ as long as $\bar d^2=100$ is
fulfilled. We therefore use the simple prescription $d_i=10$ for all
$i=1\ldots n$. 

In the bimodal case only the likelihood differs which is given
by 
\begin{equation}
 \label{example-plpp-bimodal}
  p_l^b(x)= \frac{1}{21} p_l^u(x) + \frac{20}{21}p_l^u(-x) 
\end{equation} 
and hence consists of two Gaussians with the same variance centered at
$x$ and $-x$ respectively. It is important, however, that their
relative weights are markedly different from each other. Our special
choice of parameters implies that in equilibrium the region around
$-x$ should be sampled $20$ times as often as the one around $x$. 

With the choices for prior and likelihood given above the
prior-predictive value as defined in (\ref{eq:defppv}) can be
calculated analytically. We get for both cases the result 
\begin{equation} 
 \label{eq:example-ppv}
 P(d|\mathcal{M})
 =(2\pi(\sigma_l^2+\sigma_p^2))^{-n/2}
  \exp( -\frac{\Vert d \Vert^2 }{2(\sigma_l^2+\sigma_p^2)})\; .
\end{equation}
With the parameter values chosen this yields $\ln P(d|\cM)=-476.358$. 

In the following we will test thermodynamic integration and our
procedure against the exact result. In order to present a fair
comparison between the performances of the numerical methods 
each simulation will comprise the same number ($10^9$) of MC
steps. This number will, however, be divided in different ways between
the number $M$ of intermediate $\beta$-values, the number $N$ of MC
steps per $\beta$-value for thermodynamic integration, (the number $N$ of MC
steps per run for the exponential average) , and the number $N_c$ of runs to
estimate the distribution of $R$, (to estimate the distribution of 
$\ln P_{\mathrm{td}}$).

The thermodynamic integration scheme (\ref{eq:thdint}) requires estimates 
of $\lan \llik\ran_{\beta_m}$ at appropriately chosen values $\beta_m$
of $\beta$. As a rule $\lan \llik\ran_\beta$ is a smooth function
of $\beta$ and few such values will be sufficient. Trajectories
$\{x_t\}$ are generated for each $\beta_m$ by the standard 
Metropolis algorithm. First a starting point $x_1$ is chosen at
random, e.g. from the prior distribution $p_p$ or taken directly from the 
endpoints at the previous $\beta_m$. Subsequent moves for
$t=2,...,(N-1)$ are obtained by generating a trial step $x_t \to x'$
from the distribution  
\begin{equation}
  \label{eq:trial}
  (2\pi \sigma_{\mathrm{step}}(\beta)^2)^{-n/2} \;
      \exp \big(-\frac{\Vert x' - x_t \Vert^2}
  {2 \sigma_{\mathrm{step}}(\beta)^2}   \big)
\end{equation}
The step is accepted, $x_{t+1}=x'$, with probability 
\begin{equation}
  \label{eq:pacc}
  p_{\mathrm{accept}}=\min[1,P_{\beta_m}(x')/P_{\beta_m}(x_t)]\, .
\end{equation}
and rejected, $x_{t+1}=x_{t}$, otherwise. The choice
$\sigma_{\mathrm{step}}(\beta) =0.25 ({1}/{\sigma_p^2} +
{\beta}/{\sigma_l^2})^{-1/2}$ ensures a good acceptance ratio for all
$\beta$ since it adapts to the width of $P_{\beta}(x)$ as given in
(\ref{eq:defPbeta}). The first steps in the trajectories $\{x_t\}$
are not yet characteristic for the equilibrium distribution. We
therefore discard the first $60\%$ of steps for equilibration. The rest
is thinned out by discarding all but every tenth step to suppress
correlations. From the remaining values the average 
$\langle \ln p_l(x)\rangle_\beta$ is determined. We then calculate
$\ln P(d|\cM)$ by integrating a cubic spline interpolation of
(\ref{eq:thdint}). The procedure is repeated $N_c$ times to obtain an 
average of the log prior-predictive value, $\ln P_{\mathrm{td}}$,
together with an error estimate.

\begin{figure}[htbp]
 \begin{center}
  \includegraphics[width=8cm]{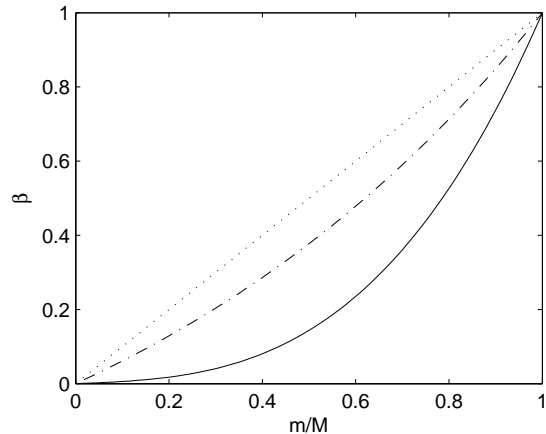}
 \end{center}
\caption{Plot of the three protocols used in the simulations. The
  dotted line shows the linear protocol (\ref{eq:lin}), the
  dashed-dotted one the exponential protocol (\ref{eq:exp}), and the
  full one the polynomial protocol defined by (\ref{eq:poly}). }
\label{fig:betaprotocols}
\end{figure}

For the simulation of (\ref{eq:rescont}) we have first to define the
protocol $\beta(t)$ according to which the parameter $\beta$ will be
changed from 0 to 1. We will use three different protocols with
$t_f=1$. Introducing $M$ equidistant time points   
\begin{equation*}
    t_m = m\cdot \frac{t_f}{M} \; , \quad m=1,...,M 
\end{equation*}
these protocols are defined respectively by
(cf.~fig.\ref{fig:betaprotocols})  
\begin{align}
\beta^{\mathrm{lin}}_m&=\frac{m}{M} \label{eq:lin}\\
\beta^{\mathrm{poly}}_m&=0.05(\frac{m}{M}) +0.95(\frac{m}{M})^3  
     \label{eq:poly} \\
\beta^{\mathrm{exp}}_m&=\frac{e^{m/M}-1}{e-1}\; .\label{eq:exp}
\end{align}

The trajectories $\{x_t\}$ are generated by the Metropolis algorithm
in a similar way as in the simulation of thermodynamic integration,
with, however, a few crucial differences. First, the number $M$ of
intermediate $\beta$-values is much higher now. 
Second, $\beta$ and therefore the acceptance
probability (\ref{eq:pacc}) changes along the trajectory.
Third, the starting point $x_1$ {\it must} now be sampled from the
prior $\pr$. Fourth, no equilibration is necessary and hence no points
will be discarded. 

At each moment when $\beta$ changes a new contribution is added to $R$
according to (\ref{eq:defR}). After $N_c$ trajectories have been
simulated, a histogram of $R$-values is generated from which the 
average $\langle R\rangle$, the variance $\sigma_R$ and 
the exponential average $\ln\left\langle e^R \right\rangle$ together
with an error estimate are calculated. 

\section{Results}

\subsection{Unimodal likelihood}
\begin{figure}[t]
 \begin{center}
  \includegraphics[width=8.5cm]{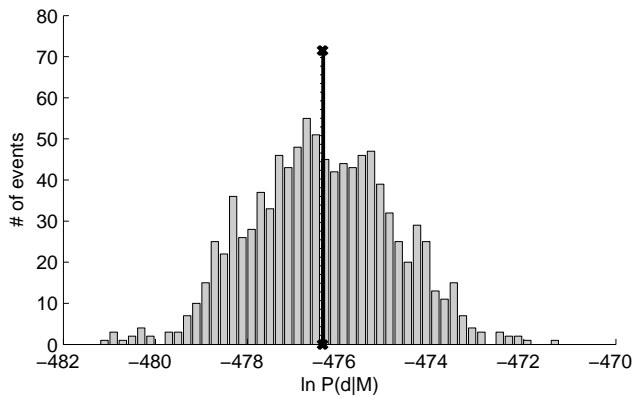}
 \end{center}
 \caption{ Histogram of the logarithm of the prior-predictive value
   for the unimodal case as
   obtained from thermodynamic integration using $M=100$ intermediate
   $\beta$-values determined according to (\ref{eq:poly}), $N=10000$
   MC steps per $\beta$ value and $N_c=1000$ runs. The dotted vertical
   line indicates the average $\ln P_{\mathrm{td}}=-476.386\pm 0.05$,
   the full one the exact result $\ln P=-476.358$.} 
 \label{fig:D7}
\end{figure}

\begin{figure}[htbp]
 \begin{center}
  \includegraphics[width=8.5cm]{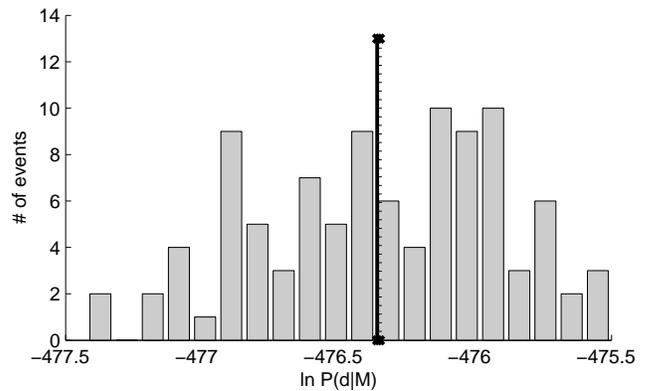}
 \end{center}
 \caption{Same as fig.\ref{fig:D7} with $M=20$, $N=500000$, and
   $N_c=100$ resulting in 
   $\ln P_{\mathrm{td}}=-476.349\pm 0.04$. } 
 \label{fig:D9}
\end{figure}

Representative results for the unimodal model from thermodynamic
integration simulations are shown in figs.\ref{fig:D7} and
\ref{fig:D9} as histograms for $\ln P(d|\cM)$ together with their
averages and the exact result (\ref{eq:example-ppv}). The intermediate
values of $\beta$ where chosen according to (\ref{eq:poly}) but this
is not very crucial. As can be seen a very good estimate of the
prior-predictive value may be obtained. 

From (\ref{eq:defPbeta}) we infer that the intermediate distributions
$P_{\beta}(x)$ are all Gaussians in this case and equilibration is
hence no problem. This is also corroborated by the comparison between
figs.\ref{fig:D7} and \ref{fig:D9} which show that a few very long
trajectories do not yield substantially better results than many long
trajectories. Accordingly thermodynamic integration works well. 

\begin{figure}[htbp]
 \begin{center}
  \includegraphics[width=8.5cm]{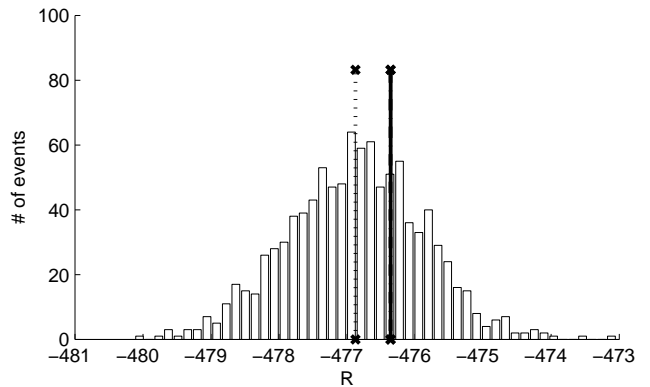}
 \end{center}
\caption{Histogram of $R$-values for the unimodal case as obtained by
  simulating   (\ref{eq:rescont}) for the polynomial protocol
  (\ref{eq:poly}) using 
  $N_c=1000$ trajectories with $N=M=10^6$ steps each. The vertical
  lines show the mean $\langle R\rangle=-476.876$ (dotted), the 
  estimate $\ln\langle e^R\rangle=-476.362 \pm 0.05$ (dashed-dotted), and
  the exact result $\ln P=-476.358$ (full). The variance of the
  histogram is given by $\sigma_R=1.0$ .}
\label{fig:D2}
\end{figure}

\begin{figure}[htbp]
 \begin{center}
  \includegraphics[width=8.5cm]{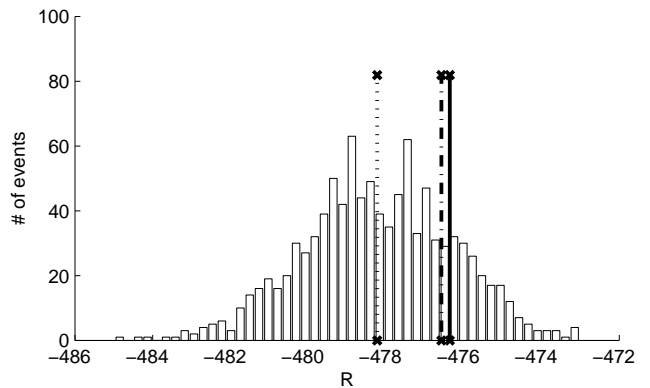}
 \end{center}
\caption{Same as fig.\ref{fig:D2} for the exponential protocol 
  (\ref{eq:exp}). In this case $\langle R \rangle=-478.226$,
  $\sigma_R=1.9$, and $\ln\langle e^R\rangle =-476.574 \pm 0.09$.}
\label{fig:D4}
\end{figure}

\begin{figure}[htbp]
 \begin{center}
  \includegraphics[width=8.5cm]{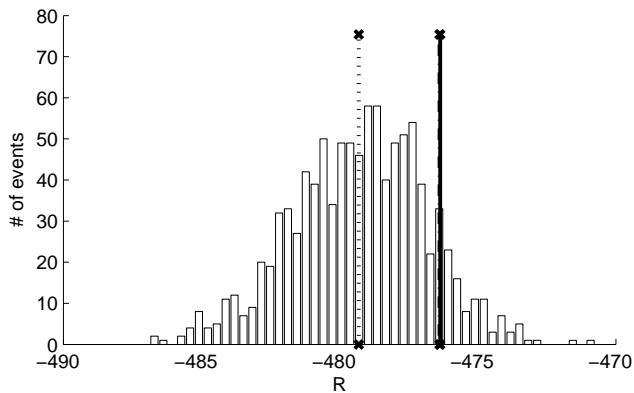}
 \end{center}
\caption{Same as fig.\ref{fig:D2} for the linear protocol
  (\ref{eq:lin}) yielding $\langle R \rangle=-479.308$, $\sigma_R=2.4$,
  and $\ln\langle e^R\rangle=-476.384 \pm 0.3$.}
\label{fig:D13}
\end{figure}

\begin{figure}[htbp]
 \begin{center}
  \includegraphics[width=8.5cm]{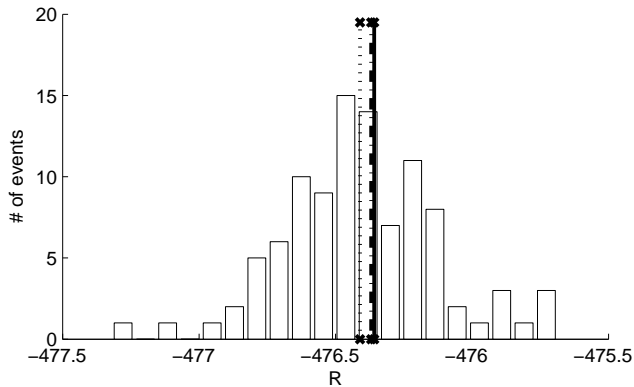}
 \end{center}
 \caption{Same as fig.\ref{fig:D2} using again the polynomial
   protocol (\ref{eq:poly}) but now with $N_c=100$ runs consisting of
   $M=N=10^7$ steps each. The results are now 
   $\langle R\rangle=-476.41$, $\sigma_R=0.3$, and $\ln\langle
   e^R\rangle =-476.370 \pm 0.03$.}
 \label{fig:D5}
\end{figure}

Results from simulations of (\ref{eq:rescont}) for the unimodal case
are shown in figs.\ref{fig:D2}-\ref{fig:D5}. 

Figs.\ref{fig:D2}, \ref{fig:D4}, and \ref{fig:D13} highlight the 
influence of the protocol $\beta(t)$, all other parameters are the
same. As can be seen the $R$-distributions produced are characterized by
different mean values $\langle R\rangle$ and different widths
$\sigma_R$. For $\beta^{\mathrm{poly}}(t)$ we get the narrowest and for 
$\beta^{\mathrm{lin}}(t)$ the widest distribution. Nevertheless the
estimate $\ln \langle e^R\rangle$ barely changes and is for all three
cases rather near to the exact value. Lower values of $\langle R 
\rangle$ and larger ones for $\sigma_R$ compensate each other
(cf. (\ref{eq:cum})) leaving the estimate for the prior-predictive 
value almost the same. Still, as far as the error in the
estimate is concerned, a narrow distribution of $R$ is advantageous
and correspondingly $\beta^{\mathrm{poly}}(t)$ performs best. 

Making the trajectories longer, decreases the variance in $R$
further as can be seen in fig.\ref{fig:D5} but leaves less 
realizations $N_c$ for the exponential average $\langle e^R\rangle$.
For the present case, however, the estimate for the prior-predictive
value remains reliable. 

In the simulations we have observed that our method gives best results
for many intermediate values of $\beta$. Therefore we have chosen for
$M$ the maximal possible number, $M=N$, implying that $\beta$ is
changed after {\it each} MC-step. From the discussion around
eq.(\ref{eq:limittdint}) this means that we are using our method in a
regime where it is very similar to thermodynamic integration. This
makes sense: for simple situations without equilibration problems
thermodynamics integration works fine and our more general method
yields comparable results for protocols which are near to a
quasi-static process. 

\subsection{Bimodal likelihood}
\begin{figure}[htbp]
 \begin{center}
  \includegraphics[width=8.5cm]{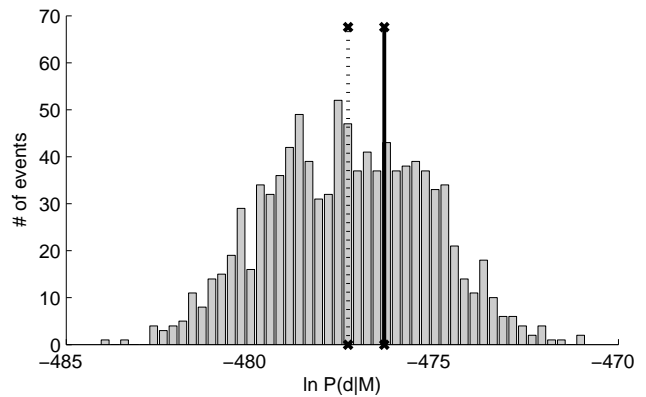}
 \end{center}
 \caption{Histogram of the logarithm of the prior-predictive value
   for the bimodal case as obtained from thermodynamic integration. 
   Parameters and meaning of the lines are the same as in 
   fig.(\ref{fig:D7}). The result is now 
   $\ln P_{\mathrm{td}}=-477.34\pm 0.07$ the exact value is still 
   $\ln P=-476.358$}   
 \label{fig:bimodal1_Nt100_NR1000_NMC10000_poly}
\end{figure}

\begin{figure}[htbp]
 \begin{center}
  \includegraphics[width=8.5cm]{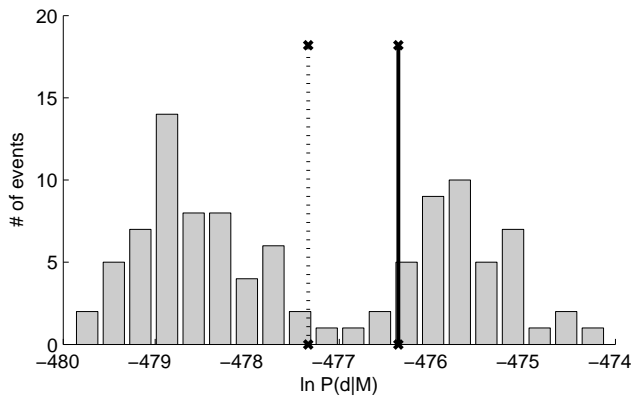}
 \end{center}
 \caption{Same as fig.\ref{fig:bimodal1_Nt100_NR1000_NMC10000_poly}
   with parameters as in fig.\ref{fig:D9}. The result is 
   $\ln P_{\mathrm{td}}=-477.338\pm 0.16$. } 
 \label{fig:bimodal2_Np128_Nt20_NR100_NMC500000_Nthin1_poly}
\end{figure}

The results obtained from thermodynamic integration for the bimodal
case are displayed in
figs. \ref{fig:bimodal1_Nt100_NR1000_NMC10000_poly} and 
 \ref{fig:bimodal2_Np128_Nt20_NR100_NMC500000_Nthin1_poly}. 
It is clearly seen that the good performance of the unimodal case is
not reached. The estimate for the prior-predictive value differs
substantially from the exact value. This failure may be traced back to
the incomplete equilibration between the two maxima of the
likelihood. In the beginning of the simulation starting with the prior
which is symmetric around $x=0$ the regions around $x=-1$ and $x=1$
are populated by the MC trajectories with roughly the same
density. Later in the simulation transitions between the regions are
extremely rare and consequently the different prefactors in 
(\ref{example-plpp-bimodal}) are not properly reproduced. It is this
incomplete equilibration which is typical for multimodal distributions
that impede a satisfactory performance of thermodynamic integration. 
As shown in
fig.\ref{fig:bimodal2_Np128_Nt20_NR100_NMC500000_Nthin1_poly} this
failure cannot be mitigated by using longer trajectories since the
equilibration over the barrier at $x=0$ is simply too slow. 

\begin{figure}[htbp]
 \begin{center}
  \includegraphics[width=8.5cm]{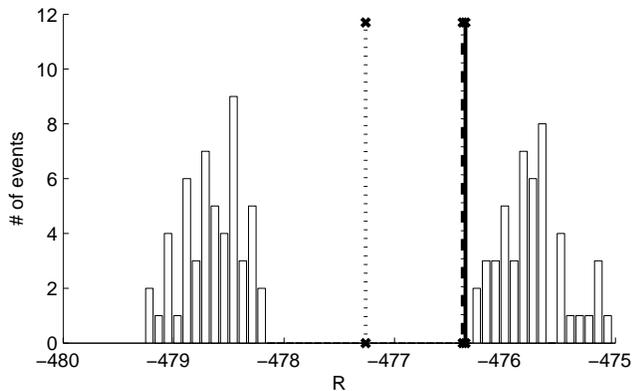}
 \end{center}
 \caption{Histogram of $R$-values for the bimodal case as obtained by
  simulating (\ref{eq:rescont}) for the polynomial protocol
  (\ref{eq:poly}) using $N_c=100$ runs consisting of $M=10^6$
  $\beta$-steps and $N=10^7$ steps altogether. The results are 
   $\langle R\rangle=-477.26$, $\sigma_R=0.15$, and $\ln\langle
   e^R\rangle =-476.38 \pm 0.10$.} 
 \label{fig:bimodal2_Np128_Nt1000000_NR100_NMC10_Nthin1_poly}
\end{figure}

\begin{figure}[htbp]
 \begin{center}
  \includegraphics[width=8.5cm]{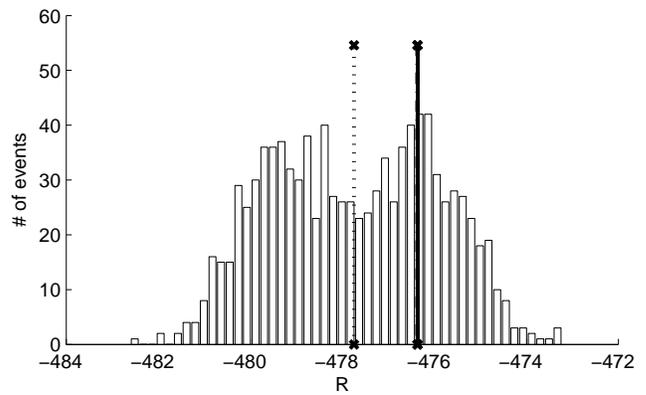}
 \end{center}
 \caption{Same as
   fig.\ref{fig:bimodal2_Np128_Nt1000000_NR100_NMC10_Nthin1_poly} with
   $N_c=1000$ runs consisting of $M=10^5$ $\beta$-steps and $N=10^6$
   steps all together resulting in 
   $\langle R\rangle=-477.74$, $\sigma_R=0.06$, and $\ln\langle
   e^R\rangle =-476.371 \pm 0.06$.} 
 \label{fig:bimodal1_Nt100000_NR1000_NMC10_poly}
\end{figure}

Results for the simulation of (\ref{eq:rescont}) for the bimodal case
are displayed in
figs.\ref{fig:bimodal2_Np128_Nt1000000_NR100_NMC10_Nthin1_poly} and
\ref{fig:bimodal1_Nt100000_NR1000_NMC10_poly}. 
As can be seen the accuracy of the estimates for the prior-predictive
value are much better than those from thermodynamic integration. In
fact the quality of the results is comparable to those for the
unimodal case. This may seem surprising since the bimodal structure of
the histogram of $R$ clearly indicates that the realizations from the MC
simulation are again captured by one of the two maxima of the
likelihood. However, the weight factor $e^R$ differs for the two
subsets of trajectories in exactly such a way as to produce the
correct prefactors in front of the two parts of the posterior
distribution, cf. (\ref{eq:crooks}). As a consequence a precise
estimate of the prior-predictive value can be obtained although no final
equilibration was reached. In this situation our method is hence
superior to thermodynamic integration. 

\section{Discussion}

In the present note we have introduced a new method to numerically
determine the prior-predictive value in a Bayesian inference problem
from MC simulations. Our method derives from a variant of the
Jarzynski equation \cite{Jar} 
\begin{equation}
  \label{eq:JE}
  \langle e^{-\beta W}\rangle=e^{-\beta \Delta F}
\end{equation}
that allows to determine the free energy difference $\Delta F$ between
two {\it equilibrium} states of a system at inverse temperature
$\beta$ from an {\it exponential average} of the work $W$ done in a 
{\it non-equilibrium} transition between the two states. In
statistical mechanics this equation has been used already to find
differences in free energy from fast-growth simulations \cite{HeJa}. 

The method proposed in the present paper incorporates two approaches
as limiting cases that are used already to determine the
prior-predictive value, namely straight MC estimation and
thermodynamic integration. The method was shown to work well in a
simple unimodal example in which its efficiency was comparable with
thermodynamic integration. It proved to be superior in the bimodal
example. 
%The simulations presented are intended to
%demonstrate that the new method indeed works and shows a performance
%comparable with thermodynamic integration. 
Our numerical implementation of both algorithms is not optimal.
The amount of samples discarded in thermodynamic integration certainly
can be reduced and the protocol $\beta^{\mathrm{poly}}$ for our procedure
is not very sophisticated.

However, the example chosen with Gaussians for both
prior and likelihood is rather remote from real applications so that a
fine-tuning of the procedures for this special case seems to be
somewhat ill-advised. A comparison of the methods when applied to a more
realistic setup with the complications alluded to in the introduction
and when implemented in a more optimal way is left for future work. 

The main advantage of the new method presumably lies in its applicability to 
multimodal systems that resist naive equilibration approaches, and its great
flexibility parametrized by a {\it protocol function} $\beta(t)$ which
may be adapted to the particular problem under consideration. We 
therefore hope that the method will provide a useful extension of the
box of tools available to perform model selection in the framework of Bayesian
data analysis. 

\vspace*{.5cm}

{\bf Acknowledgment:} We are indebted to Prof.~Dr.~Volker Dose for many
stimulating discussions.

\end{document}